# Arbitrary Choice of Basic Variables in Density Functional Theory. I. Formalism

Masahiko Higuchi *

*Institute of Solid State and Materials Research, P.O. Box 270016, D-01171 Dresden, Germany*

Katsuhiko Higuchi

*Department of Electrical Engineering, Hiroshima University, Higashi-Hiroshima 739-8527, Japan*

The Hohenberg-Kohn theorem of the density functional theory is extended by modifying the Levy constrained-search formulation. The new theorem allows us to choose arbitrary physical quantities as the basic variables which determine the ground-state properties of the system. Moreover, the theorem establishes a minimum principle with respect to variations in the chosen basic variables as well as with respect to variations in the density. By using this theorem, the self-consistent single-particle equations are derived. $N$ single-particle orbitals introduced reproduce the basic variables. The validity of the theory is confirmed by the examples where the spin-density or paramagnetic current-density is chosen as one of the basic variables. The resulting single-particle equations coincide with the Kohn-Sham equations of the spin-density functional theory (SDFT) or current-density functional theory (CDFT), respectively. By choosing basic variables appropriate to the system, the present theory can describe the ground-state properties more efficiently than the conventional DFT.

**I. Introduction**

The density functional theory (DFT)[1,2] provides the most powerful method to calculate the ground-state properties of materials. The Hohenberg-Kohn (HK) theorem[1] laid the foundation for the DFT. The theorem states that the electron density $\rho(r)$ determines the ground-state wave function and all other ground-state properties of the many-body system, whereby $\rho(r)$ is regarded as the basic variable in the DFT. The theorem also gives the variational principle with respect to the electron density. The practical scheme for calculating ground-state properties was provided by Kohn and Sham.[2] They introduced the noninteracting fictitious system and successfully derived the single-particle equations with the aid of the HK theorem. They are called the Kohn-Sham (KS) equations. The electron density can be reproduced correctly by means of the KS orbitals in the fictitious system.

The DFT has been extended to the suitable density functional frameworks by treating the characteristic quantities as the basic variables. For example, in the spin density functional theory (SDFT)[3,4] and its relativistic theory (RSDFT),[5-8] the spin density $m(r)$ and $\rho(r)$ are chosen as the basic variables which determine the ground-state of the spin-polarized system. The SDFT has the following merits in comparison with the original DFT. One is that we can get the ground-state values of both the electron density and spin density, while the DFT reproduces the electron density alone. Another merit is concerned with the simplicity of the approximate form of the exchange-correlation energy functional $E_{xc}[\rho, m]$. Due to the explicit treatment of the spin density as the basic variable, we can construct the simpler exchange-correlation energy functional than the conventional DFT.[9]

Another simple example of the extended DFT is found in the current-density functional theory (CDFT)[10-14] and its relativistic extension, the relativistic current- and spin-density functional theory (RCSDFT).[15-19] In the CDFT, the paramagnetic current density $j_p(r)$ is chosen as a basic variable as well as $\rho(r)$. The CDFT also has the merits which are analogous to those of the SDFT. Namely, not only can one get the paramagnetic current density reproduced by the KS orbitals, but the simple exchange-correlation energy functional $E_{xc}[\rho, j_p]$ can sufficiently describe the effects that would require a highly complicated functional $E_{xc}[\rho]$ in the original DFT.

In order to enjoy the above merits in general cases, it is essential to choose as the basic variables the quantities which characterize the ground-state properties of the system. Such characteristic quantities differ in individual systems. Therefore, we need the extended HK theorem which holds for arbitrarily chosen basic variables. In this paper, we shall develop the generalization of the Levy constrained-search formulation[20-22] so as to get such a theorem.

The constrained-search formulation provides the correspondence between the basic variables and the wave function irrespective of the external fields and potentials. So far, the constrained-search formulation has been applied to several cases by some authors. Electron density and off-diagonal elements of the density matrix were treated as the basic variables in both works of Levy[22] and Percus.[23] Perdew and Zunger have used the constrained-search procedure to construct the rigorous framework of the SDFT.[24] Erhard and Gross have employed the constrained-search approach, and derived the sum rules of the exchange-correlation energy functional of the CDFT.[25] In order to overcome the symmetry dilemma of the KS theory, the constrained-search formulation has been extended to the symmetrized one by Theophilou[26] and Görling,[27, 28] and further discussed by Theophilou[29] and Katriel *et al.*[30] The constrained-search approaches for the excited states have been developed by Görling,[31] Levy and Nagy,[32] Nagy and Levy.[33] Thus, the constrained-search formulation has been pursued in the specific cases. The purpose of this paper is to give the theoretical framework, in which arbitrary characteristic quantities of the system can be chosen as the basic variables, by utilizing the constrained-search procedure. Due to the arbitrary choice of the basic variables, the merits above illustrated by the SDFT and CDFT are maintained in the present theory.

The organization of this paper is as follows. In Sec. II, we present the extension of the HK theorem. The extended theorem guarantees that the ground-state wave function is determined by the basic variables which are chosen appropriately to the system. It is also shown that there exists a minimum principle with respect to variations in the chosen basic variables as well as with respect to variations in the density. In Secs. III and IV, the self-consistent single-particle equations are derived on the basis of the extended HK theorem. In order to confirm the validity of the present theory, it is shown in Sec. V that the present theory can reproduce the SDFT and CDFT formulations. Finally in Sec VI, we summarize the results and give some comments on the present theory.

## II. Extension of the Hohenberg-Kohn theorem

### A. Basic variables

Let us consider the many-body system described by the non-relativistic Hamiltonian

$$\hat{H} = \hat{T} + \hat{W} + \int \hat{\rho}(r) v_{\text{ext}}(r) dr , \qquad (2\text{-}1)$$

where $v_{\text{ext}}(r)$ is an external electromagnetic potential. $\hat{T}$, $\hat{W}$ and $\hat{\rho}(r)$ are the operators of the kinetic energy, electron-electron interaction energy, and electron density, respectively. In this system, the HK theorem holds for the basic variable $\rho_0(r)$ which is defined as the expectation value of $\hat{\rho}(r)$ with respect to the ground state $|\Psi_0\rangle$, i.e., $\rho_0(r) \equiv \langle \Psi_0 | \hat{\rho}(r) | \Psi_0 \rangle$.[1,2] In the conventional constrained-search formulation,[20-22] the functional defined by

$$F[\rho] = \underset{\Psi \to \rho}{\text{Min}} \langle \Psi | \hat{T} + \hat{W} | \Psi \rangle \qquad (2\text{-}2)$$

is introduced so as to eliminate the $v$-representability problem of the original HK theorem. Concerning the existence of the minimum in Eq. (2-2), the essential property is weak lower semicontinuity of $\langle \Psi | \hat{T} + \hat{W} | \Psi \rangle$ in the Hilbert space of the $\Psi$. Then, since $\{\Psi | \|\Psi\| = 1\}$ is weak compact,[34] the minimum exists, if $\{\Psi | \Psi \mapsto \rho, \|\Psi\| = 1\}$ is weak closed (See, Theorem 3.3 in Ref. 21). Searching over all antisymmetric wave functions that yield a particular $\rho(r)$, which is denoted by $\Psi \to \rho$ in Eq. (2-2), $F[\rho]$ gives the minimum expectation value of $\hat{T} + \hat{W}$. The procedure is illustrated in Fig. 1 (a). The *N*-electron Hilbert space is divided into the subsets, in each of which all wave functions integrate to a particular $\rho(r)$. Following the diagram on page 59 in Ref. 35, a minimizing wave function for a particular $\rho(r)$ is denoted by a dot in the subset.[35] Since the constrained-search formulation guarantees the equality $\langle \Psi_0 | \hat{H} | \Psi_0 \rangle = \langle \Psi[\rho_0] | \hat{H} | \Psi[\rho_0] \rangle$, the first theorem of Hohenberg and Kohn is immediately proven.[20]

For the purpose of the extension of the HK theorem, we consider the following constrained-search;

$$F[\rho, X] \equiv \underset{\Psi \to (\rho, X)}{\mathrm{Min}} \langle \Psi | \hat{T} + \hat{W} | \Psi \rangle, \qquad (2\text{-}3)$$

where $X(r)$ is an arbitrary physical quantity which is defined uniquely, for example, the spin density $m(r)$ or the paramagnetic current density $j_\mathrm{p}(r)$. Again, the minimum exists in Eq. (2-3), if $\{\Psi | \Psi \mapsto (\rho, X), \|\Psi\| = 1\}$ is weak closed. In the following we assume that $(\rho(r), X(r))$ is of that type. This constrained-search gives a minimum expectation value of $\hat{T} + \hat{W}$ among the antisymmetric wave functions that yield both $\rho(r)$ and $X(r)$. In this case, *N*-electron Hilbert space is divided into the smaller subsets (see Fig. 1(b)), in each of which all wave functions integrate to a particular pair of $\rho(r)$ and $X(r)$. A minimizing wave function for a particular pair of $\rho(r)$ and $X(r)$ is denoted in Fig. 1(b) by a dot in the smaller subset. Since the minimum value is determined by $\rho(r)$ and $X(r)$, we can express it as the functional of $\rho(r)$ and $X(r)$ in the left-hand side of Eq. (2-3). A minimizing wave function is denoted by $\Psi[\rho, X]$.

Suppose that a ground-state of $\hat{H}$ exists, and $\rho(r)$ and $X(r)$ for the ground-state are denoted by $\rho_0(r)$ and $X_0(r)$, respectively. From the definition of $\Psi[\rho_0, X_0]$ we have

$$E_0 = \langle \Psi_0 | \hat{H} | \Psi_0 \rangle = \langle \Psi[\rho_0, X_0] | \hat{H} | \Psi[\rho_0, X_0] \rangle . \qquad (2\text{-}4)$$

This means that under the conditions given the ground-state wave function can be obtained by finding the wave functions which minimize the expectation value of $\hat{T}+\hat{W}$ and yield both $\rho_0(r)$ and $X_0(r)$. Thus, there is a correspondence between the ground-state wave function $\Psi_0$ and a pair of $\rho_0(r)$ and $X_0(r)$. We get

$$\Psi[\rho_0, X_0] = \Psi_0 . \tag{2-5}$$

It should be mentioned that the degeneracy of the ground-state does not affect the above discussion as well as in the case of the conventional constrained-search formulation.[20] If the ground-state is degenerate, all of the ground-state wave functions may be obtained by the above-mentioned procedure repeatedly.

## B. Variational principle

Suppose again that $\hat{H}$ has an $N$-particle ground-state. The Rayleigh-Ritz principle is given by the variational search of minimum energy among all antisymmetric wave functions, i.e.,

$$E_0 = \underset{\Psi}{\text{Min}} \langle \Psi | \hat{H} | \Psi \rangle . \tag{2-6}$$

The above variational search among all antisymmetric wave functions is divided into two steps. First, we consider the subset of $N$-electron Hilbert space, in which all wave functions yield a given pair of $\rho(r)$ and $X(r)$. We minimize $\langle \Psi | \hat{H} | \Psi \rangle$ in the subset;

$$\begin{aligned}\underset{\Psi \to (\rho, X)}{\text{Min}} \langle \Psi | \hat{H} | \Psi \rangle &= \underset{\Psi \to (\rho, X)}{\text{Min}} \langle \Psi | \hat{T} + \hat{W} | \Psi \rangle + \int \rho(r) v_{\text{ext}}(r) dr \\ &= F[\rho, X] + \int \rho(r) v_{\text{ext}}(r) dr,\end{aligned} \tag{2-7}$$

where we use the fact that all wave functions in the subset yield the same $\rho(r)$. As the second step,

we minimize (2-7) over all pairs of $\rho(r)$ and $X(r)$, and obtain the minimum value of $\langle \Psi | \hat{H} | \Psi \rangle$ in the $N$-electron Hilbert space:

$$
\begin{aligned}
E_0 &= \underset{\Psi}{\text{Min}} \langle \Psi | \hat{H} | \Psi \rangle \\
&= \underset{(\rho, X)}{\text{Min}} \left\{ \underset{\Psi \to (\rho, X)}{\text{Min}} \langle \Psi | \hat{H} | \Psi \rangle \,\middle|\, \int \rho \, d\mathbf{r} = N, \int \left| \nabla \rho^{1/2} \right|^2 d\mathbf{r} < \infty \right\} \\
&= \underset{(\rho, X)}{\text{Min}} \left\{ F[\rho, X] + \int \rho(\mathbf{r}) v_{\text{ext}}(\mathbf{r}) d\mathbf{r} \,\middle|\, \int \rho \, d\mathbf{r} = N, \int \left| \nabla \rho^{1/2} \right|^2 d\mathbf{r} < \infty \right\}.
\end{aligned}
\tag{2-8}
$$

Here we use the convention that $\text{Min}\{A \,|\, C\}$ means taking the minimum value of A under the condition C. We define the energy functional $E[\rho, X]$ by

$$
E[\rho, X] \equiv F[\rho, X] + \int \rho(\mathbf{r}) v_{\text{ext}}(\mathbf{r}) d\mathbf{r}, \tag{2-9}
$$

then Eq. (2-8) is rewritten as

$$
E_0 = \underset{(\rho, X)}{\text{Min}} \left\{ E[\rho, X] \,\middle|\, \int \rho \, d\mathbf{r} = N, \int \left| \nabla \rho^{1/2} \right|^2 d\mathbf{r} < \infty \right\}. \tag{2-10}
$$

We have

$$
E_0 = E[\rho_0, X_0]. \tag{2-11}
$$

Therefore, the energy functional defined by Eq. (2-9) takes the minimum value $E_0$ for the correct ground-state values of the basic variables, $\rho_0(\mathbf{r})$ and $X_0(\mathbf{r})$. In other words, we obtain the variational principle with respect to the basic variables $\rho(\mathbf{r})$ and $X(\mathbf{r})$. This principle can also be regarded as the extension of the conventional HK theorem which states the variational principle for $\rho(\mathbf{r})$.

**III. Kinetic energy functional**

In the conventional DFT, the kinetic energy functional of the noninteracting fictitious system is defined as

$$T_s[\rho] = \min_{\{\phi_i\} \to \rho} \sum_{i=1}^{N} \langle \phi_i | -\frac{\hbar^2 \nabla^2}{2m} | \phi_i \rangle, \quad \int \rho d\mathbf{r} = N, \quad \int |\nabla \rho^{1/2}|^2 d\mathbf{r} < \infty, \quad (3\text{-}1)$$

where the notation $\{\phi_i\}$ means the set of $N$ orthonormal orbitals which construct the single Slater determinant $\Phi$, and $\{\phi_i\} \to \rho$ indicates that the search is constrained among all $\{\phi_i\}$ which yield the prescribed electron density $\rho(\mathbf{r})$. $\rho(\mathbf{r})$ is given by the expectation value with respect to the single Slater determinant:

$$\rho(\mathbf{r}) = \langle \Phi | \hat{\rho}(\mathbf{r}) | \Phi \rangle = \rho[\{\phi_i(\mathbf{r})\}] = \sum_{i=1}^{N} \phi_i^+(\mathbf{r}) \phi_i(\mathbf{r}). \quad (3\text{-}2)$$

In the present framework we shall adopt the common expression for the kinetic energy of the noninteracting fictitious system;

$$T_s[\rho, X] \equiv \min_{\{\phi_i\} \to (\rho, X)} \sum_{i=1}^{N} \langle \phi_i | -\frac{\hbar^2 \nabla^2}{2m} | \phi_i \rangle, \quad \int \rho d\mathbf{r} = N, \quad \int |\nabla \rho^{1/2}|^2 d\mathbf{r} < \infty, \quad (3\text{-}3)$$

where the notations have the same meanings as Eq. (3-1). The basic variables $\rho(\mathbf{r})$ and $X(\mathbf{r})$ are also given by the expectation values with respect to the single Slater determinant. If $X(\mathbf{r})$ has an operator denoted by $\hat{X}(\mathbf{r})$, then $X(\mathbf{r})$ is generally written as

$$X(r) = \langle \Phi | \hat{X}(r) | \Phi \rangle = X[\{\phi_i(r)\}]. \tag{3-4}$$

The minimizing set $\{\phi_i\}$ in Eq. (3-3) is determined by the pair of $\rho(r)$ and $X(r)$. Thus, the minimizing $N$ orbitals are the functional of $\rho(r)$ and $X(r)$. It should be noted that the existence of the minimum in Eq. (3-1) has been proved by Lieb.[21] We again assume that the minimum exists in Eq. (3-3) in the similar way to Eq. (2-3), i.e., $(\rho(r), X(r))$ is supposed to be of that type.

The minimizing $N$ orbitals can be obtained by searching the minimum value of $\sum_{i=1}^{N} \langle \phi_i(r) | -\frac{\hbar^2 \nabla^2}{2m} | \phi_i(r) \rangle$ under the conditions that the orbitals are orthonormal and yield the given $\rho(r)$ and $X(r)$. In order to perform this constrained-search variational, we introduce Lagrange multiplier functions $\lambda(r)$ and $\mu(r)$ for the conditions that the minimizing $N$ orbitals yield $\rho(r)$ and $X(r)$, respectively. Lagrange multipliers $\varepsilon_{ij}$ have also introduced for the condition that the minimizing $N$ orbitals are orthonormal. Define $\Omega[\{\phi_i\}]$ by

$$\begin{aligned}\Omega[\{\phi_i\}] &\equiv \sum_{i=1}^{N} \int \phi_i^+(r) \left( -\frac{\hbar^2 \nabla^2}{2m} \right) \phi_i(r) dr + \int \lambda(r) \left\{ \sum_{i=1}^{N} \phi_i^+(r)\phi_i(r) - \rho(r) \right\} dr \\ &+ \int \mu(r) \{X[\{\phi_i(r)\}] - X(r)\} dr - \sum_{i,j=1}^{N} \varepsilon_{ij} \left\{ \int \phi_i^+(r)\phi_j(r) dr - \delta_{i,j} \right\} \end{aligned} \tag{3-5}$$

Then, the minimizing condition is given by

$$\sum_{i=1}^{N} \int \delta\phi_i^+(r) \left( \frac{\delta\Omega}{\delta\phi_i^+(r)} \right) dr + \sum_{i=1}^{N} \int \left( \frac{\delta\Omega}{\delta\phi_i(r)} \right) \delta\phi_i(r) dr = 0. \tag{3-6}$$

Substitution of Eq. (3-5) into Eq. (3-6) leads to a pair of equations which are necessary conditions on the minimizing orbitals:

$$-\frac{\hbar^2 \nabla^2}{2m}\phi_k(\mathbf{r}) + \lambda(\mathbf{r})\phi_k(\mathbf{r}) + \int \mu(\mathbf{r}')\left(\frac{\delta X[\{\phi_i(\mathbf{r}')\}]}{\delta \phi_k^+(\mathbf{r})}\right) d\mathbf{r}' = \sum_{j=1}^{N} \varepsilon_{kj}\phi_j(\mathbf{r}), \quad (3\text{-}7a)$$

$$-\frac{\hbar^2 \nabla^2}{2m}\phi_k^+(\mathbf{r}) + \lambda(\mathbf{r})\phi_k^+(\mathbf{r}) + \int \mu(\mathbf{r}')\left(\frac{\delta X[\{\phi_i(\mathbf{r}')\}]}{\delta \phi_k(\mathbf{r})}\right) d\mathbf{r}' = \sum_{i=1}^{N} \varepsilon_{ik}\phi_i^+(\mathbf{r}). \quad (3\text{-}7b)$$

The Lagrange multiplier functions $\lambda(\mathbf{r})$ and $\mu(\mathbf{r})$ should be determined by requiring the orbitals to yield a given pair of $\rho(\mathbf{r})$ and $X(\mathbf{r})$. Namely, $\lambda(\mathbf{r})$ and $\mu(\mathbf{r})$ are written as $\lambda(\mathbf{r}) = \lambda[\rho(\mathbf{r}), X(\mathbf{r})]$ and $\mu(\mathbf{r}) = \mu[\rho(\mathbf{r}), X(\mathbf{r})]$, respectively. If a given pair of $\rho(\mathbf{r})$ and $X(\mathbf{r})$ corresponds to the true ground-state, then Eqs. (3-7a) and (3-7b) coincide with the single-particle equation of the fictitious system which gives the ground-state basic variables correctly. Let us consider the simplest case as an example. If we choose $\rho(\mathbf{r})$ alone as a basic variable, Eqs. (3-7a) and (3-7b) are reduced to the equations with a potential $\lambda(\mathbf{r})$ which produces $\rho(\mathbf{r})$. Further if $\rho(\mathbf{r})$ is the ground-state value, $\lambda(\mathbf{r})$ is equal to the KS effective potential of the DFT. The details will be discussed in the next section.

**IV. Self-consistent single-particle equations**

Equations (3-7a) and (3-7b) are satisfied for any values of the basic variables. In this section, we consider the single-particle equation in the case where the given basic variables coincide with the ground-state values. The variational principle which is mentioned in Sec. II.B has to be applied to this problem because it provides the prescription of getting the correct ground-state values of the basic variables. First, we define the exchange-correlation energy functional $E_{xc}[\rho, X]$ by

$$F[\rho, X] = T_s[\rho, X] + U[\rho] + E_{xc}[\rho, X], \quad (4\text{-}1)$$

where $F[\rho, X]$ and $T_s[\rho, X]$ are respectively given by Eqs. (2-3) and (3-3), and $U[\rho]$ is the Hartree term. Substituting Eq. (4-1) into Eq. (2-9), we obtain

$$E[\rho, X] = T_s[\rho, X] + U[\rho] + E_{xc}[\rho, X] + \int v_{ext}(r)\rho(r)dr . \tag{4-2}$$

The variational principle guarantees that $E[\rho, X]$ has the minimum value if $\rho(r)$ and $X(r)$ are respectively equal to the ground-state values, $\rho_0(r)$ and $X_0(r)$. Thus, the minimizing condition $\delta E[\rho_0, X_0] = 0$ is rewritten by

$$\delta T_s[\rho_0, X_0] + \delta U[\rho_0] + \delta E_{xc}[\rho_0, X_0] + \int v_{ext}(r)\delta\rho(r)dr = 0 . \tag{4-3}$$

Taking the variation with respect to the basic variables in each term, we get

$$\begin{aligned}\delta E[\rho_0, X_0] = &-\int \left\{ \lambda[\rho_0(r), X_0(r)] - v_{ext}(r) - \int \frac{e^2 \rho_0(r')}{|r-r'|} dr' - \left.\frac{\delta E_{xc}[\rho, X]}{\delta \rho(r)}\right|_{\substack{\rho=\rho_0 \\ X=X_0}} \right\} \delta\rho(r)dr \\ &- \int \left\{ \mu[\rho_0(r), X_0(r)] - \left.\frac{\delta E_{xc}[\rho, X]}{\delta X(r)}\right|_{\substack{\rho=\rho_0 \\ X=X_0}} \right\} \cdot \delta X(r)dr \\ =&\, 0,\end{aligned} \tag{4-4}$$

In the calculation of the first term of Eq. (4-3), we utilize Eqs. (3-7a) and (3-7b) via the following relation:

$$\delta\phi_i(r) = \phi_i[\rho_0 + \delta\rho, X_0 + \delta X] - \phi_i[\rho_0, X_0].$$

Equation (4-4) leads to the final expressions of $\lambda[\rho_0(r), X_0(r)]$ and $\mu[\rho_0(r), X_0(r)]$:

$$\lambda[\rho_0(r), X_0(r)] = v_{ext}(r) + \int \frac{e^2 \rho_0(r')}{|r-r'|} dr' + \left.\frac{\delta E_{xc}[\rho, X]}{\delta \rho(r)}\right|_{\substack{\rho=\rho_0 \\ X=X_0}}, \tag{4-5}$$

$$\mu[\rho_0(r), X_0(r)] = \left. \frac{\delta E_{xc}[\rho, X]}{\delta X(r)} \right|_{\substack{\rho=\rho_0 \\ X=X_0}}. \qquad (4\text{-}6)$$

Equations (3-7a) and (3-7b) with (4-5) and (4-6) reproduce the correct ground-state values of basic variables via (3-2) and (3-4).

Let us consider the case where $\hat{X}(r)$ is generally denoted by

$$\hat{X}(r) = \sum_{n=1}^{N} \hat{x}(r; \tau_n). \qquad (4\text{-}7)$$

Here $\tau_n$ comprises the space coordinate, momentum operator and vector of Pauli matrix for the particle n, and $\hat{x}(r, \tau_n)$ is the single-particle Hermitian operator. For example, the spin-density and paramagnetic current-density operators are given in this form as seen in the next section. By using Eq. (4-7), the single-particle equation (3-7a) is reduced to

$$\hat{h}_s \phi_k(r) = \left\{ -\frac{\hbar^2 \nabla^2}{2m} + \lambda[\rho_0(r), X_0(r)] + \int \mu[\rho_0(r'), X_0(r')] \cdot \hat{x}(r'; \tau) dr' \right\} \phi_k(r) = \sum_{j=1}^{N} \varepsilon_{kj} \phi_j(r). \quad (4\text{-}8)$$

$\lambda[\rho_0(r), X_0(r)]$ and $\mu[\rho_0(r), X_0(r)]$ are real number and vector, respectively. Since the single-particle Hamiltonian $\hat{h}_s$ is a Hermitian operator, the above equation can be changed to the canonical form by a unitary transformation:

$$\left\{ -\frac{\hbar^2 \nabla^2}{2m} + \lambda[\rho_0(r), X_0(r)] + \int \mu[\rho_0(r'), X_0(r')] \cdot \hat{x}(r'; \tau) dr' \right\} \phi_k(r) = \varepsilon_k \phi_k(r). \qquad (4\text{-}9)$$

Equation (3-7b) is also converted to the canonical form which is equivalent to Eq. (4-9). Note that Eqs. (3-2) and (3-4) are left invariant under the unitary transformation. Therefore, Eqs. (4-9), (4-5), (4-6), (3-2) and (3-4) can be regarded as the self-consistent single-particle equations of the fictitious system.   The advantage of the present theory is that one can choose the arbitrary quantities as the basic variables in compliance with the electronic properties of a given many-body system.   Not only the electron density but also such quantities can be obtained by means of the self-consistent single-particle equations derived above.

If we choose $\rho(r)$ alone as a basic variable, it is easily shown that Eq. (4-9) coincides with that of the conventional KS theory.   The third term on the left-hand side of Eq. (4-9) does not appear in this case.   The equation has only the local potential $\lambda(r)$ which accords with the effective potential of the KS equation.

At the end of this section, we have a discussion concerning the variational principle (4-3). As mentioned in Sec. III, $\rho[\{\phi_i(r)\}]$ and $X[\{\phi_i(r)\}]$ are given by the expectation values with respect to the single Slater determinant $\Phi$.   Therefore, $U[\rho]$, $E_{xc}[\rho, X]$ and $\int v_{ext}(r)\rho(r)dr$ in Eq. (4-2) are also regarded as the functionals of $\Phi$.   Here, define the following functional of $\Phi$:

$$E[\Phi] \equiv \langle \Phi | \hat{T} | \Phi \rangle + U[\rho] + E_{xc}[\rho, X] + \int v_{ext}(r)\rho(r)dr . \qquad (4\text{-}10)$$

The description that follows refers to the variational principle of the above functional.   The set of the single Slater determinants can be divided into the subsets, in each of which the single Slater determinants yield a particular set of $\rho(r)$ and $X(r)$ via (3-2) and (3-4).   The variational search among all single Slater determinants is divided into two steps.   We get

$$\begin{aligned}\operatorname*{Min}_{\Phi} E[\Phi] &= \operatorname*{Min}_{\Phi}\left(\langle \Phi | \hat{T} | \Phi \rangle + U[\rho] + E_{xc}[\rho, X] + \int v_{ext}(r)\rho(r)dr\right) \\ &= \operatorname*{Min}_{(\rho,X)}\left\{\operatorname*{Min}_{\Phi \to (\rho,X)}\left(\langle \Phi | \hat{T} | \Phi \rangle + U[\rho] + E_{xc}[\rho, X] + \int v_{ext}(r)\rho(r)dr\right)\right\}.\end{aligned} \qquad (4\text{-}11)$$

The single Slater determinants in a particular subset yield the same $U[\rho]$, $E_{xc}[\rho, X]$ and $\int v_{ext}(\boldsymbol{r})\rho(\boldsymbol{r})\mathrm{d}\boldsymbol{r}$ due to the definition of the subset. Thus, Eq. (4-11) is rewritten by

$$\mathop{\mathrm{Min}}_{\Phi} E[\Phi] = \mathop{\mathrm{Min}}_{(\rho,X)}\left\{\mathop{\mathrm{Min}}_{\Phi\to(\rho,X)}\left(\langle\Phi|\hat{T}|\Phi\rangle\right) + U[\rho] + E_{xc}[\rho,X] + \int v_{ext}(\boldsymbol{r})\rho(\boldsymbol{r})\mathrm{d}\boldsymbol{r}\right\}.$$

The first term $\mathop{\mathrm{Min}}_{\Phi\to(\rho,X)}\left(\langle\Phi|\hat{T}|\Phi\rangle\right)$ is equal to $T_s[\rho, X]$ because of Eq. (3-3), i.e., the definition of $T_s[\rho, X]$. Consequently, by means of the variational principle (4-3), we get the following variational principle with respect to the single Slater determinant:

$$\begin{aligned}\mathop{\mathrm{Min}}_{\Phi} E[\Phi] &= \mathop{\mathrm{Min}}_{(\rho,X)}\left\{T_s[\rho,X] + U[\rho] + E_{xc}[\rho,X] + \int v_{ext}(\boldsymbol{r})\rho(\boldsymbol{r})\mathrm{d}\boldsymbol{r}\right\} \\ &= \mathop{\mathrm{Min}}_{(\rho,X)} E[\rho,X] \\ &= E_0.\end{aligned} \qquad (4\text{-}12)$$

This means that the variational principle with respect to the basic variables (4-3) is equivalent to that with respect to the single Slater determinant. In the original DFT, the variational principle with respect to the single Slater determinant has been discussed by Hajisaavs and Theophilou[36] to overcome the $v$-representability problem. The variational principle (4-12) is recognized as the generalization of their formulation.

## V. Examples

In this section, we apply the present theory to the typical case where the spin-density or paramagnetic current-density is reasonably chosen as one of the basic variables. Each case completely reproduces the SDFT or CDFT formulation.

### A. Spin-density functional theory

For describing the ground-state of the spin-polarized system, the spin density is considered reasonable as one of the basic variables, i.e., we choose $\rho(r)$ and $m(r)$ as the basic variables in the present theory.  When this is the case, the set of single-particle equations can be obtained by letting $X(r)$ equal to $m(r)$ in the above-mentioned discussion.  The spin-density operator $\hat{m}(r)$ is given by

$$\hat{m}(r) = -\beta_e \sum_{i=1}^{N} \delta(r - r_i) \sigma^i, \qquad (5\text{-}1)$$

where $\sigma^i$ denotes the vector of Pauli matrices and $\beta_e$ is the Bohr magneton.  The single-particle operator which corresponds to $\hat{x}(r'; \tau)$ in Eq. (4-9) is given by

$$\hat{x}(r', \tau) = -\beta_e \sigma \delta(r' - r). \qquad (5\text{-}2)$$

Substituting Eq. (5-2) into (4-9), the self-consistent single-particle equations are derived as follows:

$$\left\{ -\frac{\hbar^2 \nabla^2}{2m} + \lambda[\rho_0(r), m_0(r)] - \beta_e \mu[\rho_0(r), m_0(r)] \cdot \sigma \right\} \phi_k(r) = \varepsilon_k \phi_k(r) \qquad (5\text{-}3)$$

with

$$\lambda[\rho_0(r), m_0(r)] = v_{ext}(r) + \int \frac{e^2 \rho_0(r')}{|r - r'|} dr' + \left. \frac{\delta E_{xc}[\rho, m]}{\delta \rho(r)} \right|_{\substack{\rho=\rho_0 \\ m=m_0}}, \qquad (5\text{-}4)$$

$$\mu[\rho_0(r), m_0(r)] = \left. \frac{\delta E_{xc}[\rho, m]}{\delta m(r)} \right|_{\substack{\rho=\rho_0 \\ m=m_0}}, \qquad (5\text{-}5)$$

where the basic variables are

$$\rho_0(\bm{r}) = \sum_{k=1}^{N} \phi_k^+(\bm{r})\phi_k(\bm{r}), \qquad (5\text{-}6)$$

$$\bm{m}_0(\bm{r}) = -\beta_e \sum_{k=1}^{N} \phi_k^+(\bm{r})\bm{\sigma}\phi_k(\bm{r}). \qquad (5\text{-}7)$$

The set of these single-particle equations is completely equal to that of the SDFT. Thus, the present theory provides the SDFT formulation if $\rho(\bm{r})$ and $\bm{m}(\bm{r})$ are chosen as the basic variables.

It should be noted that there is a difference between the above SDFT formulation and previous developments of the spin-density constrained-search.[24, 35] In Ref. 24 and in Section 8.1 of Ref. 35, the minimization of the total energy is carried out with respect to not the basic variables but the minimizing $N$ orbitals of Eq. (3-3). Since the minimizing $N$ orbitals are given as the solutions of Eqs. (3-7a) and (3-7b), the minimization of the total energy should be carried out under this restriction on the orbitals. $\delta T_s[\rho, X]$ in Eq. (4-3) are calculated under this restriction as mentioned in the previous section, while the total energy is minimized in Refs. 24 and 35 without imposing the restriction on the orbitals.

**B. Current-density functional theory**

The paramagnetic current density $\bm{j}_p(\bm{r})$ is considered suitable as one of the basic variables for describing the electronic structure of the system, in which a spontaneous current exists like in the open-shell atoms and f-electron materials.[37-39] In such a case, the set of single-particle equations can be obtained by making $X(\bm{r})$ equal to $\bm{j}_p(\bm{r})$ in Eqs. (4-9), (4-5) and (4-6). The paramagnetic current-density operator $\hat{\bm{j}}_p(\bm{r})$ is given by

$$\hat{\bm{j}}_p(\bm{r}) = \frac{-i\hbar}{2m} \sum_{i=1}^{N} \{\delta(\bm{r}-\bm{r}_i)\nabla_i + \nabla_i \delta(\bm{r}-\bm{r}_i)\}. \qquad (5\text{-}8)$$

In this case, the single-particle operator $\hat{\bm{x}}(\bm{r}';\tau)$ in Eq. (4-9) is

$$\hat{x}(r'; \tau) = \delta(r' - r)\frac{-i\hbar\nabla}{2m} + \frac{-i\hbar\nabla}{2m}\delta(r' - r). \qquad (5\text{-}9)$$

Substituting Eq. (5-9) into (4-9), the self-consistent single-particle equations are given by

$$\left[\frac{p^2}{2m} + \lambda[\rho_0(r), j_{p0}(r)] + \frac{1}{2m}\{p \cdot \mu[\rho_0(r), j_{p0}(r)] + \mu[\rho_0(r), j_{p0}(r)] \cdot p\}\right]\phi_k(r) = \varepsilon_k \phi_k(r) \qquad (5\text{-}10)$$

with

$$\lambda[\rho_0(r), j_{p0}(r)] = v_{ext}(r) + \int \frac{e^2 \rho_0(r')}{|r - r'|} dr' + \left.\frac{\delta E_{xc}[\rho, j_p]}{\delta \rho(r)}\right|_{\substack{\rho=\rho_0 \\ j_p=j_{p0}}}, \qquad (5\text{-}11)$$

$$\mu[\rho_0(r), j_{p0}(r)] = \left.\frac{\delta E_{xc}[\rho, j_p]}{\delta j_p(r)}\right|_{\substack{\rho=\rho_0 \\ j_p=j_{p0}}}, \qquad (5\text{-}12)$$

where the basic variables are

$$\rho_0(r) = \sum_{k=1}^{N} \phi_k^+(r)\phi_k(r), \qquad (5\text{-}13)$$

$$j_{p0}(r) = \frac{-i\hbar}{2m}\sum_{k=1}^{N}\left[\phi_k^+(r)\nabla\phi_k(r) - \{\nabla\phi_k^+(r)\}\phi_k(r)\right]. \qquad (5\text{-}14)$$

These single-particle equations, (5-10), (5-11), (5-12), (5-13) and (5-14), coincide with those of the CDFT. It should be noted that the starting Hamiltonian of the many-body system, which is given by (2-1), does not contain the term associated with the external vector potential. In the $v$-representable CDFT scheme,[10, 11] the starting Hamiltonian includes the interaction of $j_p(r)$ with an external vector potential. If we apply the CDFT to the system in the absence of an external magnetic field, the external vector potential is vanished after deriving the Kohn-Sham equation. In other words, the "artificial" external vector potential is introduced in the *v*-representable CDFT scheme so that one can deal with $j_p(r)$ as the basic variable. The present theory allows us to choose $j_p(r)$ as one of the basic variables and to derive the Kohn-Sham equation without introducing such an "artificial" external vector potential.

## VI. Concluding Remarks

In this paper, we extend the Hohenberg-Kohn theorem by modifying the Levy constrained-search formulation. The new theorem allows us to choose the arbitrary physical quantities as the basic variables. By means of this theorem, we derive the self-consistent single-particle equations which reproduce the basic variables correctly. The single-particle equations can be recognized as an extension of the KS equations of the ordinary DFT scheme. In order to confirm the validity of the theory, we consider the case where the spin-density or paramagnetic current-density is chosen as one of the basic variables. Each case makes a reproduction of the SDFT or CDFT scheme completely.

Due to the arbitrary choice of the basic variables, the present theory has two advantages over the conventional DFT. In discussing the ground-state properties, it is to be desired that the quantities which characterize the system can be obtained correctly within energy-band theory. In the present theory, we can directly calculate such characteristic quantities by means of the single-particle orbitals because they can be chosen as the basic variables. This seems to be the substantial progress of the density functional theory. The second advantage is concerned with the accuracy of the approximate forms of the exchange-correlation energy functional. The basic variables reproduced in the fictitious system would be equal to the correct ground-state values if the exact exchange-correlation energy functional were known. It is desirable to get the exchange-correlation energy functional as accurately as possible. In the present theory, the accuracy is expected to be improved because the explicit form of the exchange-correlation energy functional is written in terms of the characteristic quantities of the system.[9] In the forthcoming paper,[40] we will derive the various types of the self-consistent single-particle equations by choosing the basic variables depending on the aspects of the electron correlation.

Finally, we shall give a note on the present theory. The present theory does not necessarily require the external potential terms which are coupled with any basic variables but the electron density. This is because the essential idea of our theory is the smaller division of the Hilbert space by the basic variables (See, Fig.1(b)). Therefore, the present theory is different from " {a}-functional theory " in which the basic variables and the corresponding external potentials are written in the general forms.[41] The present theory shows that arbitrary quantities, even if they do not appear in the Hamiltonian explicitly, can be chosen as the basic variables.

**Acknowledgements**

The authors acknowledge critical readings of this manuscript, and fruitful comments and discussions on the logical basis of the DFT by Helmut Eschrig. Especially, the mathematical aspects of the constrained-search formulation were discussed with him. The authors are also obliged to Akira Hasegawa for continual discussions on energy-band theory. One of the authors (M. H.) would like to express his thanks to the Alexander von Humboldt Foundation for facilitating both the stay and the research in Dresden, Germany.

Figure caption

Fig. 1.   The *N*-electron Hilbert space divided into the subsets.

(a) Each subset consists of the wave functions which integrate to a particular $\rho(r)$.   The conventional constrained-search (2-2) is performed in the subset.   A minimizing wave function is denoted by a dot in each subset.

(b) The set in Fig. 1 (a) is further divided into the smaller subsets.   Each subset consists of the wave functions which yield not only $\rho(r)$ but also $X(r)$.   A minimizing wave function is denoted by a dot in each subset.

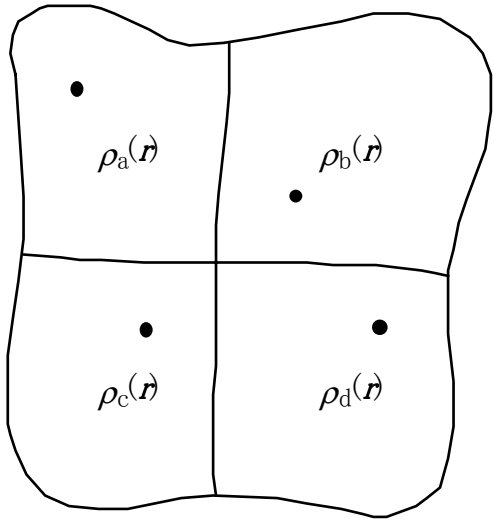 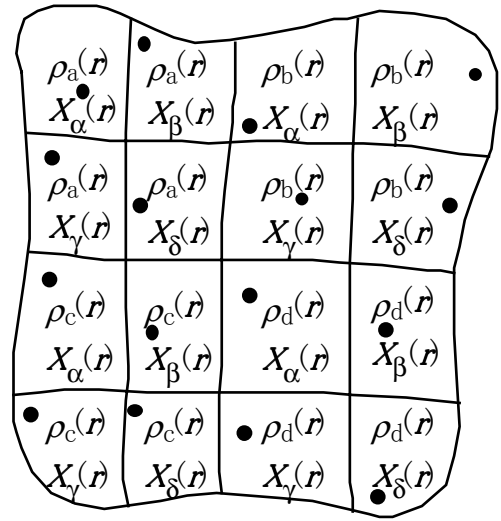

(a) (b)

Figure 1